\documentstyle[12pt,fleqn,epsf]{article}
\def\beq{\begin{equation}}
\def\eeq{\end{equation}}
\def\ct{\cos\theta}
\def\Ph{{\mit{\Phi}}}

\begin{document}

\newcommand{\2}{\otimes}
\newcommand{\9}{\rangle}
\newcommand{\6}{\langle}

\vspace*{10mm}
\begin{center}
{\large {\bf Quantum and classical descriptions of a measuring 
apparatus}}\\[15mm]

Ori Hay and Asher Peres \\[8mm]
{\sl Department of Physics, Technion---Israel Institute of Technology,
32\,000 Haifa, Israel}\\[15mm]

{\bf Abstract}\end{center}

\noindent A measuring apparatus is described by quantum mechanics while
it interacts with the quantum system under observation, and then it must
be given a classical description so that the result of the measurement
appears as objective reality. Alternatively, the apparatus may always be
treated by quantum mechanics, and be measured by a second apparatus
which has such a dual description. This article examines whether these
two different descriptions are mutually consistent. It is shown that if
the dynamical variable used in the first apparatus is represented by an
operator of the Weyl-Wigner type (for example, if it is a linear
coordinate), then the conversion from quantum to classical terminology
does not affect the final result. However, if the first apparatus
encodes the measurement in a different type of operator (e.g., the phase
operator), the two methods of calculation may give different
results.\vfill

\vfill

\newpage \begin{center}{\bf I. VON NEUMANN'S CUT}\end{center}\medskip

Quantum mechanics provides statistical predictions for the results of
measurements performed on physical systems that have been prepared in a
specified way. The preparation and measurement are performed by
macroscopic devices, and these are described in classical terms. The
necessity of using a classical terminology was emphasized by Bohr~[1],
whose insistence on a classical description was very strict. Bohr never
considered the measuring process as a dynamical interaction between an
apparatus and the system under observation. Any intermediate systems
used in that process could be treated quantum mechanically, but the {\it
final\/} instrument had a purely classical description~[2]. Measurement
was understood as a primitive notion. Bohr thereby eluded questions
which caused considerable controversy among other authors~[3,~4].

Yet, measuring apparatuses are made of the same kind of matter as
everything else, and they obey the same physical laws. It therefore
seems natural to use quantum theory in order to investigate their
behavior during a measurement. This was first attempted by von Neumann,
in his treatise on the mathematical foundations of quantum theory~[5].
In the last section of that book, as in an after\-thought, von Neumann
represented the apparatus by a single degree of freedom, whose value was
correlated to that of the dynamical variable being measured.  Such an
apparatus is not, in general, left in a definite pure state, and does
not admit a classical description. Therefore, von Neumann introduced a
second apparatus which observes the first one, and possibly a third
apparatus, and so on, until there is a final measurement, which is {\it
not\/} described by quantum dynamics and has a definite result (for
which quantum mechanics can only give statistical predictions). The
essential point that was suggested, but not proved by von Neumann, is
that the introduction of this sequence of apparatuses is irrelevant: the
final result is the same, irrespective of the location of the ``cut''
between classical and quantum physics. (At this point, von Neumann also
speculated that a final step would involve the consciousness of the
observer---a rather bizarre statement in a mathematically rigorous
monograph.)

In the present article, we introduce a dual description for the
measuring apparatus. It obeys quantum mechanics while it interacts with
the  system under observation, and then it is ``dequantized'' and is
described by a classical Liouville density, which provides the
probability distribution for the results of the measurement.
Alternatively, the apparatus may always be treated by quantum mechanics,
and be measured by a second apparatus which has such a dual description.
The question is whether these two different methods of calculation give
the same result~[6]. 

We show that a sufficient condition for agreement between the two
methods is that the dynamical variable used as a ``pointer'' by the
first apparatus be represented by an operator of the Weyl-Wigner
type~[7]. These ``quasi-classical'' operators are defined as follows:
let a classical dynamical variable, $A(q,p)$, be expressed as a Fourier
transform,

\beq A(q,p)=\int\!\!\int d\sigma\,d\tau\,
  e^{i(\sigma q+\tau p)}\,\alpha(\sigma,\tau). \eeq
Then the correponding Weyl-Wigner operator is obtained by replacing, in
the above expression, the classical variables $q$ and $p$ by the
corresponding quantum operators $\hat{q}$ and $\hat{p}$. It can be
shown that the expectation value of $\hat{A}(\hat{q},\hat{p})$ for any
quantum state, pure or mixed, is equal to the classical expression

\beq \6A\9=\int\!\!\int W(q,p)\,A(q,p)\,dq\,dp, \eeq
where $W(q,p)$ is Wigner's quasi-probability distribution~[7,~8]. If the
latter is nowhere negative, it can be interpreted as classical Liouville
distribution. In the rest of this paper, the same symbols, $q$ and $p$,
will be used for classical variables and for operators, since the
meaning of the symbol is always clear from the context and there is no
risk of confusion.

We examine two examples. In the simplest one, the pointer is described
by a linear coordinate, $q$, which is an operator of the Weyl-Wigner
type. As expected, the conversion from quantum to classical description
does not affect the final result. In the second example, the first
apparatus encodes the measurement in a phase. In that case, the operator
that we use is not of the Weyl-Wigner type, and the two methods of
calculation give different results. It is likely that the validity of
these conclusions is not restricted to the particular examples for which
we provide detailed calculations.

In both examples, the quantum system that we observe is a particle of
spin~$j$, and we want to measure the $J_z$ component. In Sect.~II, we
couple $J_z$ to the linear position, $q$, of a pointer. The latter is
then measured by a second pointer, whose linear position is $Q$. The
problem is to find the probability distribution of $Q$, for a given
initial state of the quantum system. As shown explicitly, it makes
no difference to dequantize $q$ after the first measurement, and to
always treat $Q$ classically.

In Section III, on the other hand, we couple $J_z$ to the phase,
$\theta$, of a harmonic oscillator. The second apparatus (again a linear
pointer with position $Q$) measures $\ct$, not $\theta$ itself because
the phase is not a well defined self-adjoint operator in quantum
mechanics~[9,~10]. We then find that in this case the expectation value
$\6Q\9$ is not the same when the first apparatus is treated quantum
mechanically, or classically, while it is measured by the second one.
That is, when we perform the required calculations for such a measuring
process, the result depends on the location chosen for the von Neumann
cut. Figure~1 encapsulates the difference between the two methods of
calculation.

To avoid any misunderstanding, we emphasize that the classical
description of a pointer is {\it not\/} by means of a point in phase
space, but by a Liouville density. Quantum theory makes only statistical
predictions, and any semi\-classical treatment that simulates it must
also be statistical. Our approach involves only strictly orthodox
quantum mechanics. We never speculate about modifications of the
conventional theory, such as those that have been proposed by some
authors~[4]. In particular, we do not attempt to mix classical and
quantum mechanics at any stage of the dynamical evolution~[11].

The implications of our results on the so-called ``quantum measurement
problem'' are briefly discussed in Sect.~IV. While our work may not
satisfy the desiderata of some physicists, it does prove the consistency
of those of Bohr and von Neumann, provided that the physical system that
is employed as the measuring instrument is indeed suitable for filling
that role.\bigskip

\begin{center}{\bf II. LINEAR POINTER}\end{center}

Let the system under observation be a spin $j$ particle. We want to
measure the spin component $J_z$, which satisfies, in natural units
($\hbar=1$),

\beq J_z\,|m\9=m\,|m\9, \qquad m=j, \,j-1,\ldots\ ,-j.\eeq
The initial state of the system is $\sum a_m|m\9$.

In elementary discussions of quantum measurements, there is no explicit
decription of the apparatus. The typical textbook just says that the
result of the  measurement is $m$, with probability $|a_m|^2$. The
reader may imagine a pointer, jumping from $q=0$ to $q=m$ (in suitable
units), with probability $|a_m|^2$, as a result of the measuring
process. (In the language of statistical mechanics, the Liouville
function of the pointer has peaks of size $|a_m|^2$ near $q=m$.) It is
then possible to imagine a second apparatus which measures the first
one, and has its pointer moving from $Q=0$ to $Q=q$. The readings of the
two apparatuses of course agree with each other.

In this article, we provide a quantum dynamical description for the
apparatuses. The initial state of the first pointer is specified by a
wave function $\phi(q)$. The position $q$ and its conjugate momentum,
$p=-i\partial/\partial q$, are linear operators in Hilbert space. Their
spectra extend from $-\infty$ to $\infty$. Likewise, the second
apparatus is a linear pointer with position operator $Q$, momentum
operator $P=-i\partial/\partial Q$, and initial state $\Ph(Q)$.

The joint state of the complete setup is, initially,

\beq \psi_0=\sum_m a_m\,|m\9\2\phi(q)\2\Ph(Q). \eeq
The interaction between the system and the first apparatus is
represented by the unitary operator

\beq U_1=e^{-iJ_zp}=e^{-J_z(\partial/\partial q)}. \eeq
This unitary evolution can be generated by a Hamiltonian $H_{\rm int}=
J_zp/\epsilon$, acting during a time $\epsilon$, brief enough so that
the other parts of the Hamiltonian can be neglected. However, for the
present problem, it is simpler to directly use unitary operators,
instead of exponentiating a Hamiltonian. If the state of the spin is
$|m\9$, the operator $U_1$ causes the pointer to move by $m$ length
units (with a suitable choice of units). The new state thus is, in
general,

\beq \psi_1=U_1\,\psi_0=\sum_m a_m\,|m\9\2\phi(q-m)\2\Ph(Q). \eeq

Likewise, the second pointer senses the value of $q$ and moves by $q$
units. The interaction of the two pointers is generated by

\beq U_2=e^{-iqP}=e^{-q(\partial/\partial Q)}, \label{u2}\eeq
so that

\beq \psi_2=U_2\,\psi_1=\sum_m a_m\,|m\9\2\phi(q-m)\2\Ph(Q-q). \eeq
The probability distribution of $Q$ is

\beq \int^\infty_{-\infty}dq\,\psi_2^\dagger\psi^{\phantom{\dagger}}_2=
 \int^\infty_{-\infty}dq\,\sum_m|a_m|^2\,|\phi(q-m)|^2\,|\Ph(Q-q)|^2.
 \label{convol}\eeq
This simply is the convolution of the probability distribution of the
first pointer, namely

\beq f(q)=\sum_m|a_m|^2\,|\phi(q-m)|^2,  \label{f}\eeq
with the probability distribution of the second pointer for a given
value of $q$,

\beq F(Q-q)=|\Ph(Q-q)|^2. \eeq

It will now be seen that the same result is obtained if the von Neumann
cut is placed after the first apparatus. That is, the quantum mechanical
result (\ref{f}) will be considered as a classical probability
distribution for the position of the first pointer. The initial
distribution for the second one is $F(Q)$, which is a given non-negative
function. The two pointers interact classically in such a way that

\beq f(q)\,F(Q)\to f(q)\,F(Q-q). \label{dyn}\eeq
The final result for the probability distribution of $Q$ is obviously
the same as in the quantum mechanical calculation above.

However, we still have to formally show that the postulated dynamical
evolution (\theequation) is compatible with classical mechanics. Let us
thus write $f(q)$ and $F(Q)$ as the marginals of Liouville
distributions,

\beq f(q)= \int L_1(q,p)\,dp, \label{mar1}\eeq

\beq F(Q)= \int L_2(Q,P)\,dP. \label{mar2}\eeq
The interaction of the two apparatuses lasts a very brief time,
$\epsilon$, during which the Hamiltonian is

\beq H_{\rm int}=qP/\epsilon. \label{can1}\eeq
The other parts of the Hamiltonian can be neglected. It follows that $q$
and $P$ remain constant during the measurement, and that
$\dot{p}=-P/\epsilon$ and $\dot{Q}=q/\epsilon$. When the interaction is
concluded after a time $\epsilon$, we have

\beq p\to p'=p-P, \label{can2} \eeq
and

\beq Q\to Q'=Q+q. \label{can3}\eeq
It follows that the functional form of the joint distribution evolves as

\beq L_1(q,p)\,L_2(Q,P)\to L_1(q,p+P)\,L_2(Q-q,P).\eeq
Note that the $\pm$ signs in (\theequation) are opposite to those in the
two preceding equations. This is because a Liouville distribution flows
in phase space as an incompressible fluid, and the solution of the
Liouville equation is $L'(q',p',Q',P')=L(q,p,Q,P)$.

To get the marginal distributions of $q$ and $Q$, we first integrate the
right hand side of (\theequation) over $p$, and then over $P$. The
dynamical law (\ref{dyn}) readily follows, in complete agreement with
the quantum calculation. Note that we did not have to assume any
particular form for the non-negative functions $L_1(q,p)$ and
$L_2(Q,P)$. Only the marginal probabilities (\ref{mar1}) and
(\ref{mar2}) are involved in the final result.\bigskip

\begin{center}{\bf III. ENCODING A MEASUREMENT IN A PHASE}\end{center}

We shall now measure the same quantum system with a different apparatus.
Instead of a linear pointer, we use the phase of a harmonic oscillator,
whose Hamiltonian is $H_{\rm osc}={1\over2}(p^2+q^2)$. In classical
mechanics, the phase is given by $\theta=\arctan(p/q)$. In quantum
mechanics, the issue is more complicated, as we shall see.

First, let us give, as in the preceding section, an elementary classical
description of the quantum measurement (it will later be needed for
comparison with the semi\-classical and the purely quantum treatments).
The final phase of the classical oscillator, which plays the role of a
pointer, is given by

\beq \theta=\theta_0-m\chi, \eeq
with probability $|a_m|^2$. Here, $\chi$ is any constant (we shall take
$\chi<\pi/2j$, so that there is no overlap in the final values of
$\theta$). It will be convenient to take $\theta_0=\pi/2$.

The second apparatus is, as before, a linear pointer. It is coupled to
$\ct$ (not to $\theta$ itself, for reasons that will become clear
below). The final position of the second pointer (treated classically)
thus is

\beq Q=Q_0+\ct=Q_0+\sin{m\chi}. \label{class} \eeq

This elementary classical result, for which no dynamical justification
was given, will now be compared with the one obtained by treating both
apparatuses as quantum systems.\medskip

\begin{center}{\bf A. Two quantum apparatuses}\end{center}\medskip

The first apparatus is a harmonic oscillator (e.g., one of the modes of
an electro\-magnetic field in a cavity), initially prepared in a
coherent state~[12],
 
\beq |\alpha\9=e^{-r^2/2}\sum_{k=0}^\infty (\alpha^k/\sqrt{k!})\,|k\9,
 \label{cohst} \eeq
where $\alpha$ is a complex number. On the right hand side of
(\theequation), the orthonormal basis $|k\9$ consists of eigenstates
of $H_{\rm osc}$,

\beq H_{\rm osc}\,|k\9=(k+\mbox{$1\over2$})\,|k\9, \label{nbst} \eeq
and

\beq  r^2\equiv|\alpha|^2=
  \6\alpha|H_{\rm osc}|\alpha\9-\mbox{$1\over2$}\,. \eeq
The coherent states (\ref{cohst}) minimize the uncertainty product
$\Delta p\,\Delta q=\hbar/2$, and therefore give results as close as
possible to classical physics, in particular when $r\gg1$.

The second apparatus is, as before, a linear pointer prepared in a state
$\Ph(Q)$. The joint state of the complete setup thus is, initially

\beq \psi_0=|m\9\2|\alpha\9\2\Ph(Q). \eeq
Here, we have assumed for simplicity that the quantum system is in one
of the eigenstates $|m\9$ of $J_z$ (the goal of the measurement is to
determine $m$). It is obvious that any linear combination $\sum a_m|m\9$
would give, after the quantum system is traced out, a statistical
mixture with weights $|a_m|^2$, as in Eq.~(\ref{f}).

The interaction between the system and the first apparatus is
represented by the unitary operator

\beq U_1=e^{-i\chi J_zH_{\rm osc}}. \eeq
As before, it is easy to write an interaction Hamiltonian that generates
this unitary evolution. In the present case, where $m$ has a definite
value, we can replace in (\theequation) $J_z$ by $m$. It then follows
from (\ref{cohst}) and (\ref{nbst}) that

\beq \psi_1=U_1\,\psi_0=
 e^{-im\chi/2}\,|m\9\2|e^{-im\chi}\alpha\9\2\Ph(Q). \eeq
From this point, we can safely ignore the spin state $|m\9$, since we
shall not observe again the quantum system itself, and of course we
ignore the phase factor $e^{-im\chi/2}$.

If we could now measure the phase of the parameter $e^{-im\chi}\alpha$
in the coherent state on the right hand side of (\theequation), this
would readily give us the value of $m$. This is of course impossible,
because coherent states are not mutually orthogonal~[12] and they cannot
be distinguished with certainty. At most, we may get probabilistic
indications for the value of $m$. Moreover, there is no self-adjoint
phase operator~[9]. It is however possible to define a self-adjoint
operator $C$, which is a legitimate quantum analogue of the classical
variable $\ct$. (Most authors simply call that operator $\ct$, or
$\widehat{\ct}$, instead of $C$ as we do here to avoid ambiguities.) The
reader who is not interested in computational details may skip from here
to Eq.~(\ref{skip}).

The spectrum of $C$ runs from $-1$ to 1, and it is convenient to label
the eigenvalues by $\ct$, with $0\leq\theta\leq\pi$.  The eigenstates of
$C$ are given, in terms of the number states $|n\9$, by~[9]

\beq |\ct\9=\sqrt{2/\pi}\,\sum_{n=0}^\infty \sin{[(n+1)\theta]}\,
 |n\9. \label{eigv}\eeq
They have a delta-function normalization,

\beq \6\ct|\ct'\9=\delta(\theta-\theta'),\label{ortho}\eeq
and a completeness property,

\beq \int_0^\pi|\ct\9\,\6\ct|\,d\theta={\bf 1}, \label{compl}\eeq
where {\bf 1} is the unit operator.

The interaction between the first and the second apparatuses is given,
as in (\ref{u2}), by

\beq U_2=e^{-iCP}=e^{-C(\partial/\partial Q)}. \eeq
To see how this acts on $\psi_1$, we have to expand
$|e^{-im\chi}\alpha\9$ into a sum of eigenstates of $C$. For such an
eigenstate, the evolution is

\beq U_2\,|\ct\9\2\Ph(Q)=|\ct\9\2\Ph(Q-\ct). \label{U2}\eeq
By virtue of (\ref{compl}), we have

\beq |e^{-im\chi}\alpha\9=\int_0^\pi
 d\theta\,|\ct\9\,\6\ct|e^{-im\chi}\alpha\9. \label{psi1} \eeq
The expression $\6\ct|e^{-im\chi}\alpha\9$ can be evaluated explicitly
thanks to (\ref{cohst}) and (\ref{eigv}). For brevity, let us write

\beq e^{-im\chi}\alpha=e^{i\mu}r, \eeq
where $\mu=\mu_0-m\chi$. It will be convenient to take as the initial
phase $\mu_0=\pi/2$.

The next step is to compute $\psi_2=U_2\psi_1$. Collecting all the
relevant expressions, we obtain from (\ref{U2}),

\beq |\psi_2(Q)\9=\sqrt{2\over\pi} e^{-r^2/2}\int_0^\pi d\theta\,
 \sum_{n=0}^\infty\sin[(n+1)\theta]\,\frac{r^n e^{in\mu}}{\sqrt{n!}}\,
 |\ct\9\2\Ph(Q-\ct),\eeq
where we have used a mixed notation, as in the preceding equations: the
Dirac symbol $|\enskip\9$ is used for the states of the first apparatus,
and ordinary functions of $Q$ for the second apparatus. With these
notations, the probability distribution for $Q$, irrespective of the
value of $\ct$, is given by the diagonal elements of the partly traced
density matrix: ${\rm Tr}_\theta[|\psi_2(Q)\9\6\psi_2(Q)|]$. The result
is, thanks to the orthogonality relation (\ref{ortho}),

\beq {2\over\pi} e^{-r^2}\int_0^\pi \!d\theta
 \sum_{n=0}^\infty\sin[(n+1)\theta]\,\frac{r^n e^{in\mu}}{\sqrt{n!}}\,
 \sum_{s=0}^\infty\sin[(s+1)\theta]\,\frac{r^s e^{-is\mu}}{\sqrt{s!}}\,
 |\Ph(Q-\ct)|^2.\eeq
This expression is a convolution, just as in Eq.~(\ref{convol}). It is
difficult to evaluate it explicitly, but the mean value, $\6Q\9$ , can
easily be obtained. Keeping the integration over $\theta$ for the end,
we have

\beq \int_{-\infty}^\infty Q\,dQ\,|\Ph(Q-\ct)|^2=\6Q\9_0+\ct, \eeq
and therefore

\beq \6Q\9=\6Q\9_0+\6C\9, \eeq
where

\beq \6C\9={2\over\pi} e^{-r^2}\int_0^\pi d\theta\,\ct
 \sum_{n,s=0}^\infty\sin[(n+1)\theta]\,\sin[(s+1)\theta]\,
  \frac{r^{n+s} e^{i(n-s)\mu}}{\sqrt{n!}\,\sqrt{s!}}, \eeq
is the expectation value of $C$ in the state $|e^{-im\chi}\alpha\9$
whose expansion was given in Eq.~(\ref{psi1}). We now make use of

\beq \int_0^\pi d\theta\,\ct \sin[(n+1)\theta]\,\sin[(s+1)\theta]
 \equiv \pi(\delta_{n,s+1}+\delta_{s,n+1})/4.\eeq
This gives, after some rearrangement,

\beq \6Q\9-\6Q\9_0=\cos\mu\,e^{-r^2}\,\sum_{n=0}^\infty
 r^{2n+1}\Bigm/\sqrt{n!(n+1)!}\,. \label{skip} \eeq
The coefficient, $\cos\mu\equiv\sin{m\chi}$, is the classical result
(\ref{class}) for the displacement of $Q$. The quantum motion of the
first apparatus reduces the average value of this displacement by a
factor $S(r)$, which depends on the amplitude of the coherent state in
which the oscillator was prepared:

\beq S(r)=e^{-r^2}\,\sum_{n=0}^\infty r^{2n+1}\Bigm/\sqrt{n!(n+1)!}\,.
 \label{damp1}\eeq
For small $r$, we have $S(r)\to r$. For large $r$, we note that the
ratio of consecutive terms in the infinite series is
$r^2/\sqrt{n(n+1)}$. Consecutive terms first increase, and then they
diminish and converge to zero. The main contribution to $S(r)$ comes
from the largest terms in this sum. These occur around $r\simeq n$,
where the
fraction in (\theequation) is approximately equal to $r^{2n}/n!$.
Therefore the series tends to $e^{r^2}$, and $S(r)\to1$. This is the
expected result, since a harmonic oscillator in a coherent state with
large $r$ is almost classical. Figure~1 shows a plot of the function
$S(r)$.\medskip

\begin{center}{\bf B. Semiclassical description}\end{center}\medskip

The above results will now be compared with a semiclassical treatment
similar to the one that was introduced in the preceding section. The
second apparatus is always described by classical statistical mechanics.
It is prepared in a Liouville distribution $L_2(Q,P)$, and it interacts
with the first apparatus, for which we also assume a Liouville
distribution. The latter is initially identical to the Wigner function
$W(q,p)$~[8] that results from the first stage of the measurement. It
is indeed possible to identify these two distributions, because the
first apparatus is in a coherent state, so that its Wigner function is
everywhere positive. (If we had chosen another state, whose Wigner
function had negative regions, it would have been necessary to smooth
the oscillations of $W(q,p)$, so as to make it everywhere
positive~[13].)

We must now construct an interaction between the two apparatuses in such
a way that $Q$ moves to a new value, $Q+C$, as in Eq.~(\ref{can3}). To
respect classical mechanics, this has to be a continuous canonical
transformation, generated by a Hamitonian

\beq H_{\rm int}=CP/\epsilon, \label{Hint}\eeq
as in Eq.~(\ref{can1}). Here, 

\beq C=\ct=q\Bigm/\sqrt{p^2+q^2}. \label{C} \eeq
The reader who is not interested in computational details may skip from
here to Eq.~(\ref{damp2}).

The variable canonically conjugate to $C$ is

\beq p_C=(p^2+q^2)^{3/2}/2p=H_{\rm osc}/\sin\theta, \eeq
as may be checked by computing their Poisson bracket, $[C,p_C]=1$. We
may also write $p_C$ as

\beq p_C=\pm H_{\rm osc}\Bigm/\sqrt{1-C^2}. \eeq
Note that $-1\leq C\leq1$, but for any given $C$ the domain of $p_C$
extends from $-\infty$ to $\infty$.

With the interaction (\ref{Hint}), $C$ and $P$ are constant, while 

\beq p_C\to p_C'=p_C-P,\eeq
and

\beq Q\to Q'=Q+C,\eeq
as in Eqs.~(\ref{can2}) and (\ref{can3}). The joint distribution thus
evolves as

\beq L_1(C,p_C)\,L_2(Q,P)\to L_1(C,p_C+P)\,L_2(Q-C,P).\eeq

To obtain the probability distribution of $Q$, we have to integrate the
right hand side of (\theequation) over all the other canonical
variables. First, we note that since $p_C$ extends from $-\infty$ to
$\infty$, a shift by the parameter $P$ makes no difference in the
integral: we can replace in the integrand $L_1(C,p_C-P)$ by
$L_1(C,p_C)$. This allows us to return to the original canonical
variables,

\beq L_1(C,p_C)\,dC\,dp_C=W(q,p)\,dq\,dp. \eeq
Once this is done, the integration over $P$ yields

\beq \int L_2(Q-C,P)\,dP=F(Q-C), \eeq
where $C$ is given by the right hand side of (\ref{C}), and $F(Q)$ is
the initial marginal distribution for $Q$.

As in the preceding full quantum treatment, we shall calculate the
average final value $\6Q\9$, for a given initial distribution $F(Q)$:

\beq \6Q\9=
 \int^\infty_{-\infty}Q\,dQ\int\!\!\int dq\,dp\,W(q,p)\,F(Q-C). \eeq
We again shift the origin, $Q\to Q+C$, and obtain

\beq \6Q\9=
  \6Q\9_0+\int\!\!\int dq\,dp\,W(q,p)\,\frac{q}{\sqrt{q^2+p^2}},
 \label{Wigner} \eeq
where we have replaced $C$ by its explicit value (\ref{C}), and made use
of $\int\!\int W(q,p)\,dqdp=1$ and $\int F(Q)\,dQ=1$.

Explicitly, for the coherent state $|e^{i\mu}r\9$, we have (see
ref.~[6], pp.~316 and 325),

\beq W(q,p)=\pi^{-1}\,e^{-(q-\6q\9)^2-(p-\6p\9)^2}, \eeq
where

\beq \6q\9=\sqrt{2}\,r\,\cos\mu, \eeq
\beq \6p\9=\sqrt{2}\,r\,\sin\mu.\eeq
We likewise replace $q$ and $p$ by polar coordinates (whose physical
meaning is that of action-angle variables~[9]),

\beq q=\sqrt{2}\,s\,\cos\phi,\eeq 
\beq p=\sqrt{2}\,s\,\sin\phi,\eeq
so that $dq\,dp=2\,sds\,d\phi$. The mean displacement of $Q$,

\beq \delta Q=\6Q\9-\6Q\9_{0},\eeq
is then found to be, after some rearrangement,

\beq \delta Q={2\over\pi}\int_0^\infty sds\oint d\phi
\,\cos\phi\,e^{-2s^2-2r^2+4rs\cos{(\phi-\mu)}}. \eeq
Thanks to the periodicity of $\phi$, it is possible to shift its origin 
by $\mu$, so that

\beq \delta Q={2\over\pi}\int_0^\infty sds\oint d\phi
\cos{(\phi+\mu)} e^{-2s^2-2r^2+4rs\cos\phi}. \eeq
In the expression
$\cos{(\mu+\phi)}\equiv\cos\mu\cos\phi-\sin\mu\sin\phi$, the second
term is odd in $\phi$ and does not contribute to the integral.

Since now only $\cos\phi$ is involved in the integrand, it is convenient
to remap the $s\phi$ plane so that $-\infty<s<\infty$ and
$0\leq\phi\leq\pi$. We thus obtain

\beq \delta Q={2\over\pi}\,\cos\mu\,e^{-2r^2}\int_{-\infty}^\infty sds
 \int_0^\pi d\phi\,\cos\phi\,e^{-2s^2+4rs\cos\phi}. \eeq
The exponent can be written as $-2(s-r\cos\phi)^2+2r^2\cos^2\phi$. We
shift the origin of $s$ by $r\cos\phi$, and perform the integration over
$s$ explicitly, with result:

\beq \delta Q={2\over\sqrt{\pi}}\,\cos\mu\,re^{-2r^2} \int_0^\pi d\phi\,
\cos^2\phi\,e^{2r^2\cos^2\phi}. \eeq
We then substitute $\phi=y/2$ and $\cos^2\phi=(1+\cos y)/2$, and obtain

\beq \delta Q={\cos\mu\over\sqrt{2\pi}}\,re^{-r^2}
 \int_0^\pi e^{2r^2\cos y}\,(1+\cos y)\,dy. \eeq
Finally, thanks to the identity~[14]
\beq \int_0^\pi e^{z\cos y}\,(\cos n y)\,dy\equiv\pi\,I_n(z), \eeq
we have

\beq \delta Q=\cos\mu\,\sqrt{\pi/2}\,re^{-r^2}\,[I_0(r^2)+I_1(r^2)].
 \label{damp2}\eeq
The expression that multiplies $\cos\mu$ (which was the classical
result) tends to $r\sqrt{\pi/2}$ when
$r$ is small, and to 1 when $r$ is large. It is plotted in Fig.~1.

Why is this result different from the preceding one, in
Eq.~(\ref{damp1})? The reason is that the two classically equivalent
expressions for $C$ in Eq.~(\ref{C}) are not equivalent when these
expressions become operators in quantum mechanics.  The semi\-classical
result (\theequation) was obtained by using the Wigner function $W(q,p)$
in Eq.~(\ref{Wigner}) as if it were a classical probability density.
This would be justified if the operator $C$, whose expectation value we
seek, were of the Weyl-Wigner form~[7]. However $C$, which is defined by
its spectrum and eigenstates in Eq.~(\ref{eigv}), is not of that form.
It is therefore not surprising that the semi\-classical approximation
gives a final result which is different from the quantum prediction. (On
the other hand, the linear operator $q$ that was used in Sect.~II has
the Weyl-Wigner form, and therefore the two methods of calculation
agree.)\bigskip

\begin{center}{\bf IV. SUMMARY AND OUTLOOK}\end{center}

The reader who expected to find in this article a solution of the
so-called ``quantum measurement problem'' may be disappointed. Indeed,
that problem is ill defined, and it is understood in different ways by
various authors~[3,~4]. Our way of formulating the problem---for which
we can indicate a solution---simply is to say:  quantum theorists
describe the physical world by means of a complex Hilbert space (vectors
and operators) that defies any realistic interpretation, while
experimenters find plain numbers. The experimenters manipulate measuring
instruments made of ordinary matter, for which quantum theory is assumed
valid, but the ultimate outcome of the measuring process is essentially
classical~[1,~2]. Therefore, at some stage, a transition has to be made
from the quantum formalism to a classical language.

In this article, we have shown that if the measuring apparatus is
suitably chosen (as in Sect.~II), the transition from quantum mechanics
to classical {\it statistical\/} mechanics can proceed in a consistent
way. However, as shown in Sect.~III, a ``bad'' choice of apparatus is
incompatible with a classical description (more precisely, the
semi\-classical results do not coincide with those predicted by quantum
theory, though they may asymptotically be the same for large $r$).

This brings us to the unavoidable fundamental question: what are the
properties that are necessary for a physical system to be a legitimate
measuring apparatus? Our results indicate that if an apparatus uses as
its ``pointer'' a dynamical variable represented by an operator of the
Weyl-Wigner form, it is legitimate to dequantize it and to proceed as if
its Wigner function were a classical probability density. For other
types of operators, the transition from quantum to classical mechanics
usually is only an approximation (which may be excellent if the quantum
state of the apparatus is quasi-classical).

Furthermore, the replacement of Wigner's function $W(q,p)$ by a Liouville
function $L(q,p)$ is consistent  only if $W(q,p)\ge0$. We did not check
this condition in Sect.~II, because we did not need $W(q,p)$: only the
marginal probability distribution for $q$ was required. In general, if
Wigner's function is explicitly needed, it has to be non-negative for a
semi-classical treatment to proceed. Fortunately, this condition is
likely to be fulfilled for any macroscopic apparatus which is not in a
pure state, but rather in a mixed state with $\Delta q\Delta p\gg\hbar$
(this inequality is the hallmark of being ``macro\-scopic'')~[15]. All
the negative parts of $W$ are then washed away by the coarseness of the
apparatus.

In summary, there is nothing mysterious in the transition from the
quantum world to the classical one. There is no need of invoking
anthropo\-morphic concepts, such as consciousness. Plain orthodox
quantum mechanics and classical statistical mechanics correctly
reproduce all statistical predictions that can be verified in
experiments, provided that the measuring apparatus satisfies suitable
conditions, such as those discussed above. If enough care is exercised,
no inconsistency shall arise.

\medskip\begin{center}{\bf Acknowledgments}\end{center}

OH was supported by a grant from the Technion Graduate School. Work by
AP was supported by the Gerard Swope Fund, and the Fund for
Encouragement of Research.\medskip

\begin{center}{\bf References}\end{center}

\frenchspacing
\begin{enumerate}

\item N. Bohr, in {\it Albert Einstein, Philosopher-Scientist\/} ed. by
P.~A.~Schilpp (Library of Living Philosophers, Evanston, 1949)
pp.~201--241: ``However far the phenomena transcend the scope of
classical physical explanation, \ldots\ the account of the experimental
arrangement and the results of the observations must be expressed in
unambiguous language, with the terminology of classical physics.''

\item N. Bohr, in {\it New Theories in Physics\/} (International
Institute of International Cooperation, Paris, 1939) pp.~11--45: ``In
the system to which the quantum mechanical formalism is applied, it is
of course possible to include any intermediate auxiliary agency employed
in the measuring process [but] some ultimate measuring instruments must
always be described entirely on classical lines, and consequently kept
outside the system subject to quantum mechanical treatment.''

\item J. A. Wheeler and W. H. Zurek (eds.), {\it Quantum Theory and
Measurement\/}, (Princeton University Press, Princeton, 1983).

\item J. Bub, {\it Interpreting the Quantum World\/} (Cambridge
University Press, Cambridge, 1997).

\item J. von Neumann, {\it Mathematische Grundlagen der
Quantenmechanik\/}, Springer, Berlin (1932); transl.\ by R.~T.~Beyer,
{\it Mathematical Foundations of Quantum Mechanics\/}, Princeton
University Press, Princeton (1955).

\item A. Peres, {\it Quantum Theory: Concepts and Methods\/}
(Kluwer, Dordrecht, 1993) p.~376.  

\item M. Hillery, R. F. O'Connell, M. O. Scully, and E. P. Wigner, Phys.
Reports {\bf 106}, 121 (1984). See in particular pp.~132--134.

\item E. Wigner, Phys. Rev. {\bf 40}, 749 (1932).

\item P. Carruthers and M. M. Nieto, Rev. Mod. Phys. {\bf 40}, 411
(1968).

\item J. R. Torgerson and L. Mandel, Phys. Rev. Lett. {\bf 76}, 3939
(1996).

\item O. V. Prezhdo and V. V. Kisil, Phys. Rev. A {\bf 56}, 162 (1997),
and references therein.

\item R. J. Glauber, Phys. Rev. {\bf 131}, 2766 (1963).

\item K. Husimi, Proc. Phys. Math. Soc. Japan {\bf 22}, 264 (1940). 

\item M. Abramowitz and I. A. Stegun, {\it Handbook of Mathematical
Functions\/} (Dover, New York, 1972) pp.~375--377.

\item N. D. Cartwright, Physica {\bf 83A}, 210 (1976).

\end{enumerate}\nonfrenchspacing\clearpage

\vspace*{\fill}

\noindent FIG. 1. The function $S(r)$, given by Eq.~(\ref{damp1}), is
the factor by which the mean result of the quantum measurement is
reduced, with respect to the classical result. The semi\-classical
result, given by Eq.~(\ref{damp2}), is shown by the dotted curve.

\includegraphics{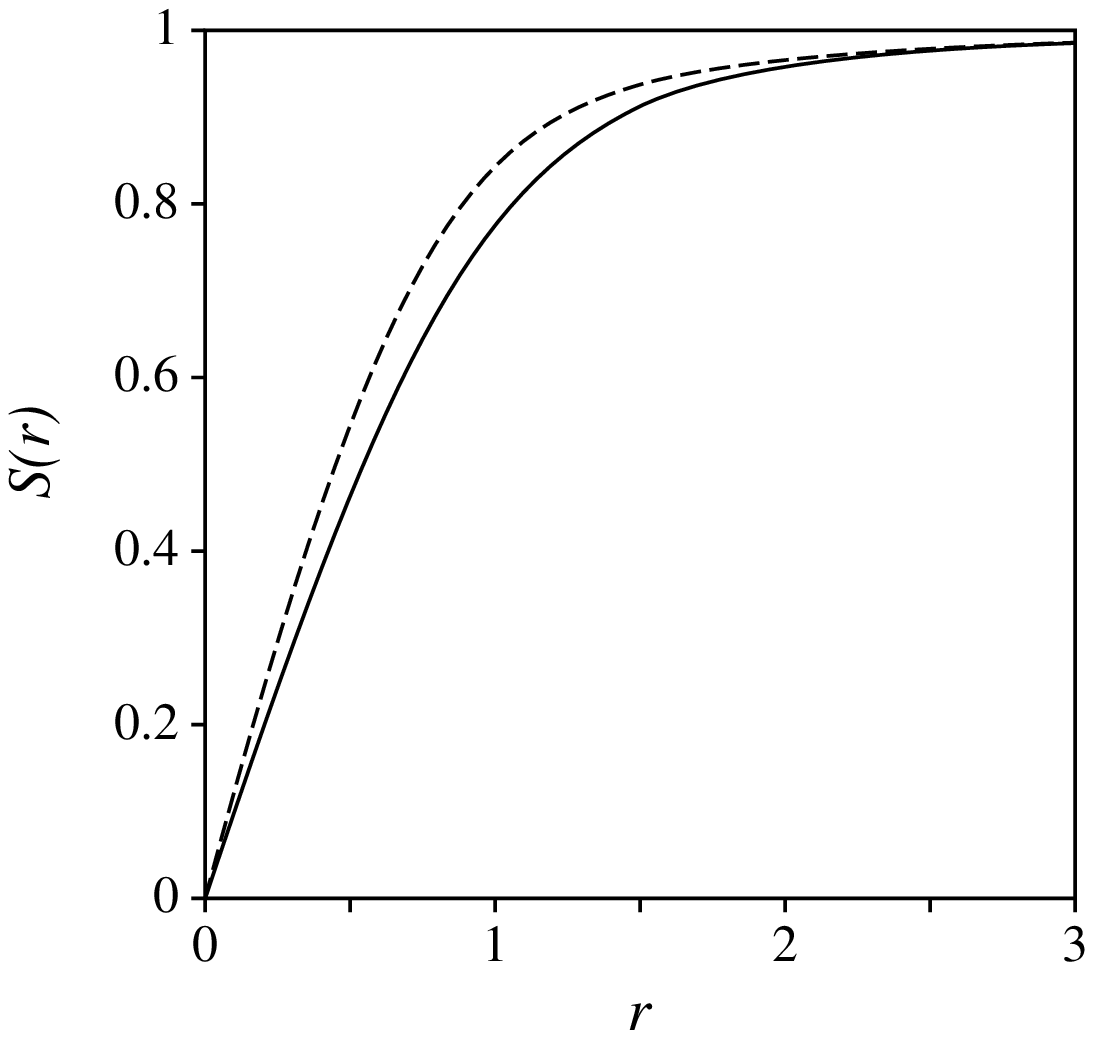}
\end{document}